\documentclass[12pt]{article}

\usepackage[T1]{fontenc}
\usepackage[utf8]{inputenc}

\usepackage{amsmath, amssymb, amsfonts, amsthm}
\usepackage{mathtools} 
\usepackage{semantic}
\usepackage{cancel}
\usepackage{slashed}

\mathlig{->}{\rightarrow}
\mathlig{|->}{\mapsto}

\usepackage{graphicx}
\usepackage{epstopdf}
\usepackage{overpic}
\usepackage{tikz}
\usetikzlibrary{decorations.pathmorphing}

\usepackage{setspace}
\usepackage{array}
\usepackage{color}
\usepackage{float}
\usepackage{textcomp}
\usepackage{comment}
\usepackage{fancyhdr}

\usepackage[
colorlinks=true,
linkcolor=blue,
urlcolor=purple,
citecolor=purple,
pdfstartview=FitV,
pdfpagemode=None,
bookmarksopen=true,
linktoc=page
]{hyperref}

\numberwithin{equation}{section}

\newcommand{\bea}{\begin{eqnarray}}
	\newcommand{\eea}{\end{eqnarray}}
\newcommand{\ba}{\begin{array}}
	\newcommand{\ea}{\end{array}}

\newenvironment{imageeq}[1][]{
	\refstepcounter{equation}
	\begin{center}
	}{
	\\	\hfill (\theequation)
	\end{center}
}

\begin{document}
\begin{flushright}
	\texttt{}
\end{flushright}

\begin{centering}
\thispagestyle{empty}

	\textbf{\Large{
Firewalls in the non-perturbative bulk Hilbert space of JT gravity }}
	
	\vspace{1.4cm}
	
	{\large  Hamed Zolfi$^{a,b}$ }
	
	\vspace{0.9cm}
	
	\begin{minipage}{.9\textwidth}\small
		\begin{center}
								{
		$^{a}$~Department of Physics, College of Science, Shiraz University, Shiraz 71454, Iran	\\
		$^{b}$~School of Particles and Accelerators,
		Institute for Research in Fundamental Sciences (IPM)
		P.O. Box 19395-5531, Tehran, Iran
	}	
	
	\vspace{0.9cm}
	{\tt  \ Emails: h.zolfi@shirazu.ac.ir}
	\\ 		
		\end{center}
	\end{minipage}
	\vspace{2cm}

	\begin{abstract}
It has been shown that a very old black hole can tunnel into a white hole through the emission of a large baby universe. This process can be modeled by a genus-one geometry corresponding to a single baby universe emission, with a tunneling probability proportional to \( t^{2} e^{-2S(E)} \), where \( t \) denotes the black hole age and \( S(E) \) its entropy at energy \( E \). The growth of this probability at late times raises the question of its behavior near \( t \sim e^{S} \). A natural possibility is that the full genus expansion, together with its non-perturbative completion, leads to saturation of the tunneling probability. Motivated by this idea, the present analysis employs a non-perturbative bulk inner product in place of the perturbative one and shows that, at late times, the probabilities of realizing firewall geometries and smooth geometries approach constant values.
	\end{abstract}
\end{centering}
\newpage
\doublespacing
\tableofcontents
\setstretch{1.1}
\setcounter{equation}{0}
\setcounter{page}{2}
\newpage
\section{Introduction}
The AdS/CFT correspondence proposes a relation between the quantum field theory partition function and the quantum gravity partition function, together with an isomorphism between the Hilbert spaces of the two theories \cite{Maldacena:1997re,Witten:1998qj}. However, longstanding difficulties remain in identifying both the partition function and the Hilbert space of the quantum theory of gravity. On the other hand, a central open question in quantum gravity is to understand the gravitational origin of the discreteness in the black hole energy spectrum \cite{Strominger:1996sh}. This discreteness controls the long-time behavior of correlation functions and the spectral form factor, and it plays a central role in other aspects of the information paradox \cite{Maldacena:2001kr, Penington:2019npb,Almheiri:2019psf,Almheiri:2019qdq,Penington:2019kki}.

Black holes are strongly chaotic systems, so their energy levels are  expected to display  the same statistical properties as the eigenvalues of a random matrix model.  Furthermore, gravitational systems--particularly black holes--exhibit non-perturbative effects. This motivates the study of eigenvalue statistics in random matrix models as a window into non-perturbative quantum gravity. 
Another concept that entered the theory of gravity some time after black holes--and indeed through the study of black holes themselves--is that of wormholes. The role of spacetime wormholes in theoretical physics remains subtle and, in many respects, conceptually confusing \cite{mal,mar,deBoer:2018ibj,DeBoer:2019yoe, Zolfi:2023bdp}.

In recent years, Jackiw--Teitelboim (JT) gravity and its supersymmetric extensions as a a simple quantum gravity theory of two dimensional negatively curved spacetimes --following the discovery of their duality with the SYK model--have provided valuable insights into low-temperature black hole dynamics and quantum chaotic behavior. These theories admit a dual description in terms of double-scaled random matrix models, which reproduce the perturbative (spacetime) topological expansion of the partition function \cite{Cotler:2016fpe,Saad:2019lba,Stanford:2019vob,Witten:2020wvy}.\footnote{For an alternative construction of the matrix model using techniques from minimal string theory, see \cite{Johnson:2019eik,Johnson:2020heh,Johnson:2020exp}.}

Given recent progress in understanding black hole microstates through the gravitational path integral, and the recognition that non-perturbative gravitational effects have considerable explanatory power, it is natural to explore whether these techniques can also clarify the interior structure of black holes
 \cite{Berkooz:2018qkz,Lin:2022rbf}.
In this context, a non-perturbative construction of the bulk Hilbert space of JT gravity has been proposed \cite{Iliesiu:2024cnh}. This framework provides a well-defined setting in which observables that probe the black hole interior can be studied beyond the perturbative regime. In particular, \cite{Iliesiu:2024cnh} considers two such operators: one that measures the interior length of the black hole, and another that measures the center-of-mass collision energy between an observer infalling from one side and a shockwave arriving from the opposite side. Both observables receive significant non-perturbative corrections at very late times, making them effective probes for potential firewall behavior in very old black holes.
It is worth noting that shortening the spatial wormhole leads to the appearance of a firewall \cite{Susskind:2015toa}. In JT gravity, the Einstein--Rosen bridge can shrink through the emission of a baby universe \cite{Saad:2019pqd}. It was further shown in \cite{Stanford:2022fdt} that a very old black hole can tunnel to a white hole by emitting a large baby universe, modeled by a genus-one geometry corresponding to a single baby-universe emission. 
The probability for tunneling to a white hole is proportional to 
\( t^{2} e^{-S(E)} \), where \( t \) is the age of the black hole and 
\( S \) and \( E \) denote its entropy and energy, respectively. 
Extensions of this analysis to an arbitrary number of baby-universe 
emissions and to setups including matter are presented in 
\cite{Zolfi:2024ldx,Zolfi:2025ieu} and \cite{Cui:2024ibh}, respectively.

The growth of the tunneling probability with $t$ raises the question of what happens when $t \sim e^{S}$. A natural possibility is that the full genus expansion, and possibly its non-perturbative completion, leads to a saturation of the tunneling probability. As mentioned above, \cite{Iliesiu:2024cnh} proposes a non-perturbative bulk inner product, whereas the analysis of Stanford and Yang \cite{Stanford:2022fdt} relies on a perturbative one. This naturally leads to the question of how the results of the latter change when revisited using the non-perturbative inner product, and in particular how the presence of geodesic boundaries (baby universes) modifies the non-perturbative inner product. To address these points, this paper is organized as follows.

 Section \ref{2} begins by establishing the JT gravity wave-function framework, while Section \ref{3} is assigned on reviewing the non-perturbative bulk Hilbert space. Section \ref{4} extends this setup to include a baby universe in non-perturbative bulk inner product. Wormhole configurations are then analyzed in two cases: without baby-universe emission (Section \ref{without}) and with baby-universe emission (Section \ref{6}). Finally, Section \ref{7} presents some remarks on the results.
\section{JT gravity wave function setup}\label{2}
The partition function of JT gravity at temperature $\beta^{-1}$ is defined by integrating over metrics on surfaces of constant negative curvature, conventionally fixed to $R = -2$. Such surfaces admit a decomposition into a trumpet geometry, with one asymptotic boundary and one geodesic boundary, together with a compact surface of genus $g$ that has multiple geodesic boundaries. This compact component is attached to the trumpet along the geodesic boundary. The partition function associated with a genus-$g$ geometry with one asymptotic boundary of length $\beta$ and $n$ geodesic boundaries, interpreted as baby universes, which is given by \cite{Saad:2019lba, Stanford:2017thb}:
\begin{figure}[H]
	\centering
	\begin{overpic}[width=0.35\textwidth]{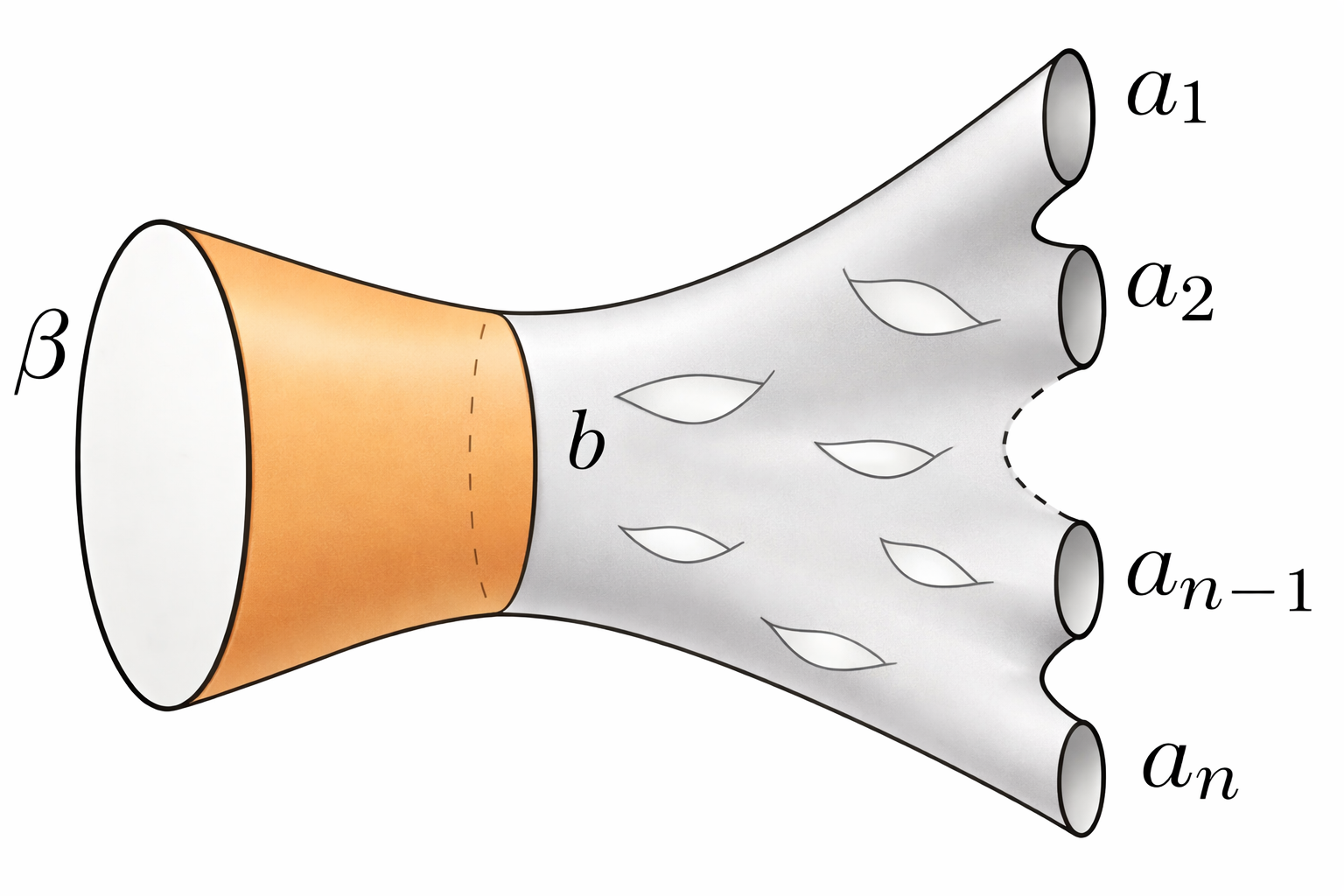}
		\put(-31,27){$\displaystyle \mathcal{Z}_{g}(\beta,\textbf{a})=$}
	\end{overpic}
	\begin{equation}\label{31}
		=	e^{S_0\chi} \int_{0}^{\infty} 
		Z_{\text{tr}}(\beta,b)\, V_{g,n+1}(b,\textbf{a}) \, b\,\mathrm{d}b.
	\end{equation}
\end{figure}
\noindent where the partition function of trumpet is:
\begin{figure}[H]
	\centering
	\begin{overpic}
		[width=0.26\textwidth]{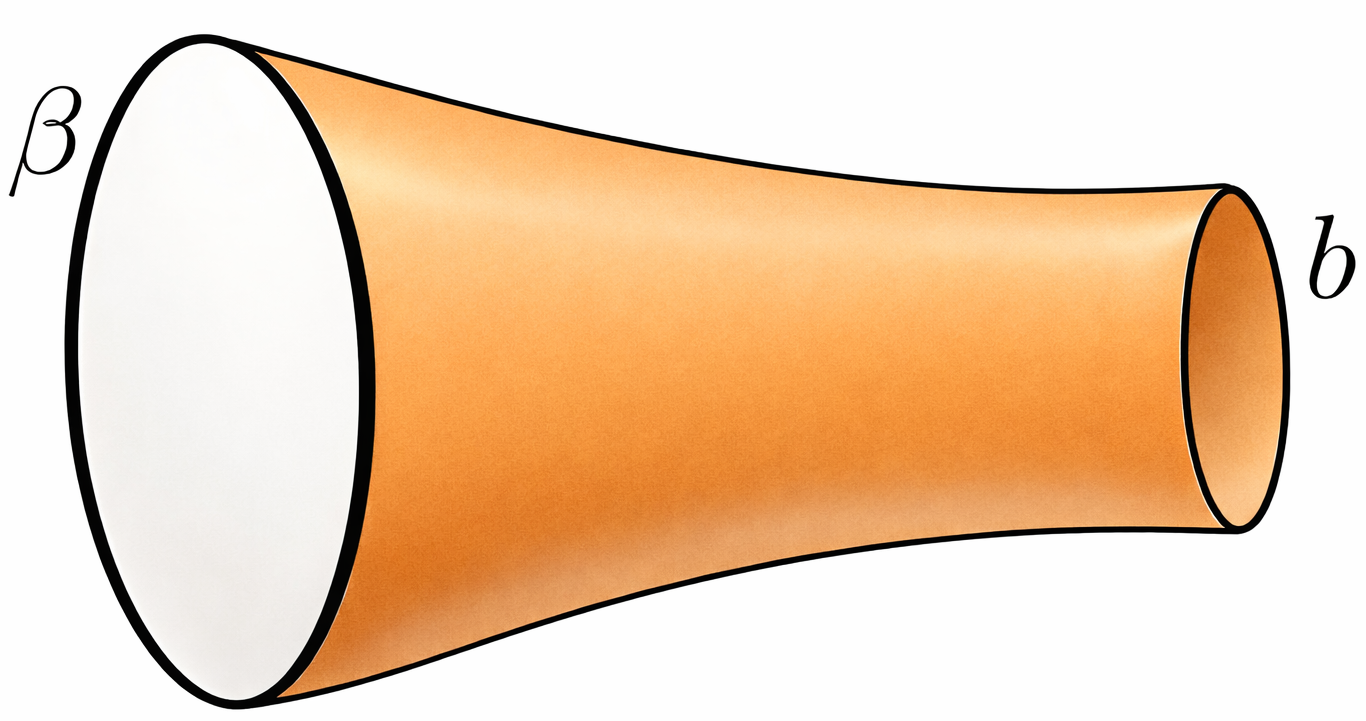}
		\put (-78,30) {$\displaystyle Z_{\text{tr}}\left( \beta,b\right)=$}			
	\end{overpic}
	\begin{equation}\label{trup}
		\hspace*{3.5cm} =\int_{0}^{\infty} \rho_{\text{tr}}\left(E,b\right)e^{-\beta E}\text{d}E=\frac{e^{-\frac{b^2}{4\beta}}}{\sqrt{4\pi \beta}},\qquad  \rho_{\text{tr}}\left(E,b\right)=\frac{\cos b\sqrt{E}}{2\pi \sqrt{E}},
	\end{equation}
\end{figure}
\noindent where $\chi$ and $S_0$ denote the Euler characteristic and ground state entropy, respectively. The function $V_{g,n+1}(b,\textbf{a})$ is the Weil--Petersson volume of the moduli space of hyperbolic Riemann surfaces of genus $g$ with $n+1$ geodesic boundaries of lengths $(b, a_1, a_2, \ldots, a_n)$ \cite{mir, mir1}. Using the analytic continuation $\beta \rightarrow \beta + \mathrm{i}(t - t')$ and an inverse Laplace transformation with respect to $\beta$, the trumpet partition function can be interpreted as a transition from an older black hole of age $t$ to a younger one with effective age $t'$, accompanied by the emission of a baby universe of size $b$ \cite{Stanford:2022fdt}. For the partition function in \eqref{31}, however, this interpretation requires restricting to a subset of the moduli space \cite{Zolfi:2025ieu}.
   
The  partition function \eqref{31} can be written differently, as follows: 
\begin{align}\label{part2}
	\mathcal{Z}_{g}	(\beta,\textbf{a}) =e^{S_0\chi }\int_{0}^{\infty} \rho_{g}\left(E,\textbf{a}\right)e^{-\beta E}\text{d}E,
\end{align}
where $\rho_{g}\left(E,\textbf{a}\right)$ denotes the eigenvalue density, with the subscript $g$ indicating the genus of the surface.
To derive the expression of $\rho_{g}\left(E,\textbf{a}\right)$, one can use the substitution of the trumpet partition function $Z_{\text{tr}}\left(\beta,b\right)$ 
into the relation \eqref{31} as follows:
\begin{equation}\label{rho}
	\mathcal{Z}_{g}	(\beta,\textbf{a}) =e^{S_0\chi }\int_{0}^{\infty} \int_{0}^{\infty}\rho_{\text{tr}}\left(E,b\right)V_{g,n+1}(b,\textbf{a})e^{-\beta E}  b\text{d}b\text{d}E.
\end{equation}
The expression for $\rho_{g}\left(E,\textbf{a}\right)$ is obtained by comparing equations \eqref{rho} and \eqref{part2}. The resulting expression is: 
\begin{equation}\label{density}
	\rho_{g}\left(E,\textbf{a}\right)=\int_{0}^{\infty}\rho_{\text{tr}}\left(E,b\right)V_{g,n+1}(b,\textbf{a})b\text{d}b.
\end{equation}

In the absence of matter, JT gravity is equivalent to a non-relativistic particle moving in a Liouville potential \cite{Bagrets:2016cdf, har}. Within this framework, the only dynamical freedom is the renormalized geodesic length of the wormhole, a gauge-invariant observable. Consequently, the partition function in \eqref{31} can be alternatively expressed as the gluing of two wave functions of Hartle-Hawking states with a specific amplitude: 
\begin{figure}[H]
	\centering
	\begin{overpic}
		[width=0.35\textwidth]{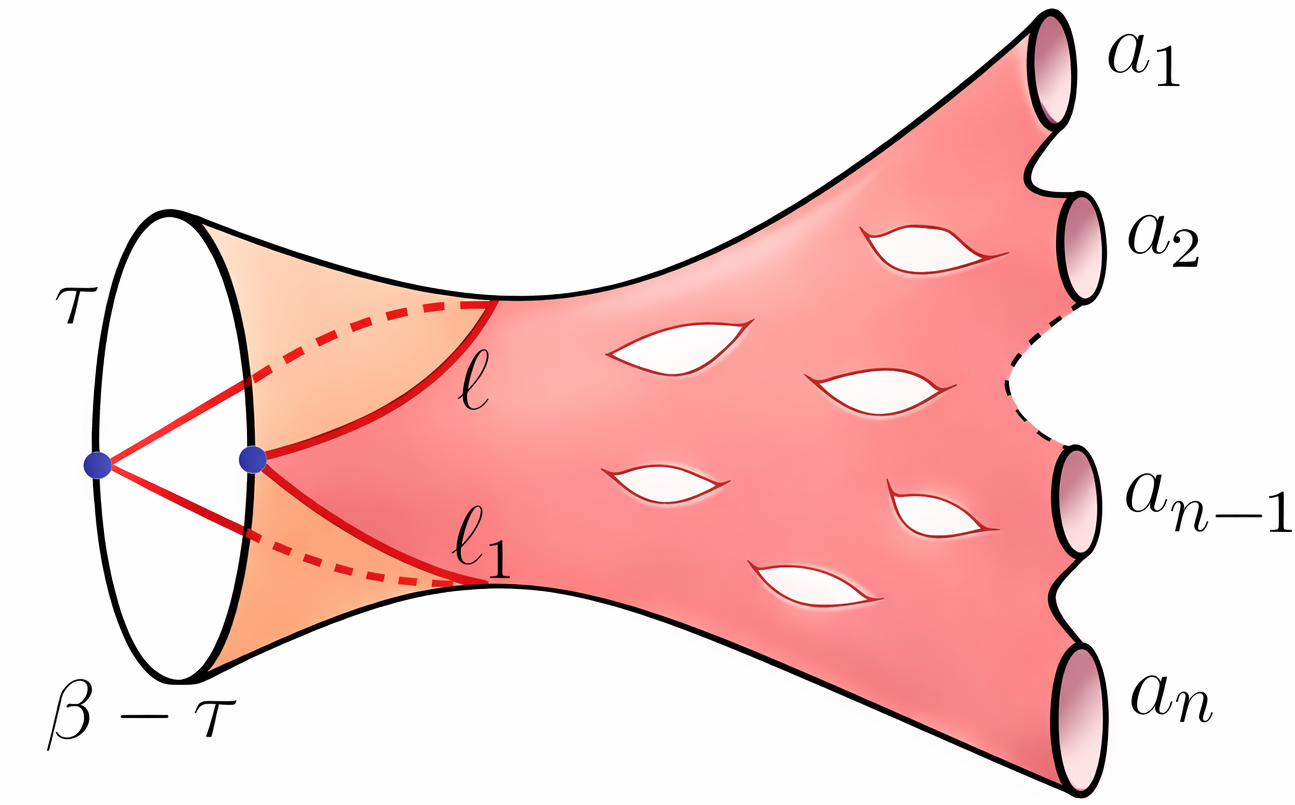}
		\put (-30,24) {$\displaystyle \mathcal{Z}_{g}	(\beta,\textbf{a})=$}		
	\end{overpic}
	\vspace*{2mm}	
	\begin{equation}\label{a1}
		=\int_{-\infty}^{\infty}\langle\tau|\ell\rangle\langle\ell,\textbf{a}|\ell_{1}\rangle_{g}\langle\ell_{1}|\beta-\tau\rangle \text{d}\ell_{1}\text{d}\ell,
	\end{equation}
\end{figure}
\noindent where the amplitude $\langle\ell,\textbf{a}|\ell_{1}\rangle_{g}$ is:
\begin{figure}[H]
	\centering
	\begin{overpic}
		[width=0.25\textwidth]{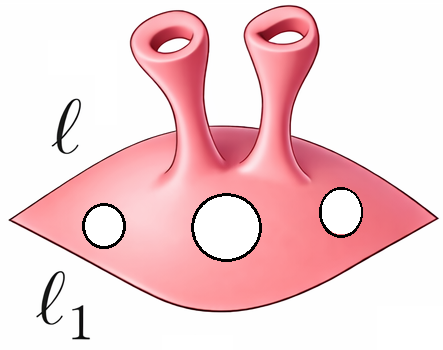}
		\put (-59,28) {$\displaystyle \langle\ell,\textbf{a}|\ell_{1}\rangle_{g}=$}		
	\end{overpic}	
	\begin{equation}\label{a}
		=e^{S_0\chi }\int_{0}^{\infty}\langle\ell|E\rangle\langle E|\ell_{1}\rangle \rho_{g}\left(E,\textbf{a}\right)\text{d}E.
	\end{equation}
\end{figure}
\noindent The expression $\langle \ell, \textbf{a} | \ell_1 \rangle_g$ represents the amplitude for a wormhole of regularized length $\ell_1$ to emit $n$ baby universes and transition into another wormhole of  length $\ell$, with the underlying geometry having genus $g$.
For latter calculations, it would be useful to insert $ \rho_{g}\left(E,\textbf{a}\right) $ from expression \eqref{density} into \eqref{a}, so one gets: 
\begin{equation}\label{amp}
	\langle\ell,\textbf{a}|\ell_{1}\rangle_{g}=e^{S_0\chi }\int_{0}^{\infty} \int_{0}^{\infty}\rho_{\text{tr}}\left(E,b\right)V_{g,n+1}(b,\textbf{a})\langle\ell|E\rangle\langle E|\ell_{1}\rangle b\text{d}b\text{d}E.
\end{equation}
The wave functions $\langle \ell \mid E \rangle$, which are energy eigenstates of the Liouville Hamiltonian, together with the disk-topology density of states $\rho_{0}(E)$, are given by \cite{har,yang,kit,Bagrets:2017pwq}:
\begin{equation}\label{d}
	\langle \ell \mid E \rangle = 2^{3/2} K_{2\mathrm{i}\sqrt{E}}\left(2 e^{-\ell/2}\right),
	\qquad
	\rho_{0}(E) = \frac{\sinh\left(2\pi \sqrt{E}\right)}{(2\pi)^{2}},
\end{equation}
which satisfy the orthogonality and completeness relations in the perturbative limit $S_0 \rightarrow \infty$:
\begin{align}
	\label{orl}
	\langle E | E' \rangle_{\text{pert}} 
	&= \int_{-\infty}^{\infty} \langle E | \ell \rangle \langle \ell | E' \rangle  \mathrm{d}\ell 
	= \frac{\delta(E - E')}{\rho_0(E)}, \\\vspace*{3mm}
	\label{2s10}
	\langle \ell | \ell' \rangle_{\text{pert}} 
	&= \int_{0}^{\infty} \langle \ell | E \rangle \langle E | \ell' \rangle \rho_0(E)  \mathrm{d}E 
	= \delta(\ell - \ell').
\end{align}
By  substituting \eqref{amp} into \eqref{a1} and integrating over $\ell$ and $\ell_1$ one  obtains \eqref{rho}.

Relation \eqref{2s10} reflects the fact that, in hyperbolic disk geometry, there exists a unique geodesic connecting two boundary points. This uniqueness implies that the lengths of the two geodesics must coincide. When $S_0$ is finite and the non-perturbative effects are included, the geodesic is no longer unique on each geometry. As a result, one might expect that a Hilbert space defined in terms of geodesic lengths is no longer well-defined. To address this issue, the proposal of \cite{Iliesiu:2024cnh} was to modify the inner product \eqref{2s10} by computing it through the path integral over all geometries bounded by a pair of geodesics of lengths $\ell$ and $\ell^{\prime}$, an approach, first presented in \cite{Gao:2021uro}.
The
next section reviews the non-perturbative Hilbert space in two-dimensional JT gravity.

\section{Review on the non-perturbative bulk Hilbert space }\label{3}
As mentioned in the introduction, JT gravity is dual to an average over an ensemble of quantum mechanical boundary theories, rather than to a single system.\footnote{For a one-dimensional extension of this duality--relating a random ensemble of two-dimensional CFTs to three-dimensional Einstein gravity--see \cite{Maloney:2020nni,Afkhami-Jeddi:2020ezh}.} Accordingly, the full quantum gravity partition function with a single asymptotic boundary of regularized length $\beta$, obtained by summing over all genera, can be written in the following form:
\begin{align}\label{r1}
	Z_{\text{JT}}\left(\beta \right)= \left\langle \text{Tr} \, e^{-\beta H} \right\rangle.
\end{align}
This equality is significant because the matrix model provides a non-perturbative definition of JT gravity and is therefore expected to enable the construction of a well-defined bulk Hilbert space beyond perturbation theory. On the right-hand side, the angle brackets denote an average over an appropriate ensemble of $L \times L$ random matrices (with $L \gg 1$),
\begin{align}\label{rmp}
	\left\langle \text{Tr} \, e^{-\beta H} \right\rangle=\frac{\int\text{d}H e^{-L \text{Tr}V\left( H\right)} \text{Tr}e^{-\beta H}}{\int\text{d}H e^{-L \text{Tr}V\left( H\right)}},
\end{align}
where $H$ is interpreted as the Hamiltonian of a dual quantum mechanical system, and $V(H)$ is a potential function.
For random matrix ensembles, the action of the symmetry group allows one to express $H$ in a canonical form in terms of the real “eigenvalues” $E_i$. For the three Wigner--Dyson ensembles, the measure for integration over the $E_i$ takes the form:
\begin{align}\label{rmm}
	\text{d}H \longrightarrow \prod_{k}\text{d}E_{k}\prod_{i<j}|E_i-E_j|^{\beta},
\end{align}
with $\beta = 2, 1$, or $4$ for GUE, GOE, or GSE, respectively.
The relevant matrix ensembles have a partition function given by \cite{Saad:2019lba,mehta2004random}: 
\begin{align}\label{zm}
	\mathrm{Z}&=\prod_{i=1}^{L}\int \exp\left(-L\sum_{i=1}^{L}V\left( E_i\right)  \right) \prod_{i<j}^{L}\left(E_i-E_j \right)^2  \text{d}E_i\nonumber\\&=\prod_{i=1}^{L}\int \exp\left(-L\sum_{i=1}^{L}V\left( E_i\right) +\sum_{i\neq j}^{L}\log\left(E_i-E_j \right) \right)  \text{d}E_i,
\end{align}
where the potential $V(E)$ determines which dilaton gravity theory is being studied.
In terms of the density of states, the  relation \eqref{r1} becomes:
\begin{align}\label{r1p}
	\left\langle \text{Tr} \, e^{-\beta H} \right\rangle
	&= \biggl\langle \int_{0}^{\infty} \sum_{i} \delta(E - E_i) e^{-\beta E} \mathrm{d}E \biggr\rangle \nonumber \\
	Z_{\text{JT}}(\beta) &= \int_{0}^{\infty} \overline{\rho(E)} e^{-\beta E} \mathrm{d}E,
\end{align}
where the function
\begin{align}\label{r2}
	\overline{\rho\left(E \right)}=\sum_{g=0}^{\infty} e^{\left(1-2g \right) S_0}\rho_{g}\left(E \right),
\end{align}
is the large $L$ limit of the density of
eigenvalues of the matrix $H$. Note that $\rho_g(E)$ is given by the relation \eqref{density} for the case with no geodesic boundary.
To derive the total density of states $\overline{\rho\left(E \right)}$, one multiplies \eqref{r1p} by $e^{\beta E}$ and integrates over $\beta$ along an inverse Laplace transform contour. According to \eqref{rmp}, since the averaging procedure involves integrals over the eigenvalues $\{ E_i \}$, the factor $e^{\beta E}$ and the $\beta$-integral can be brought inside the ensemble average. More explicitly:
\begin{align}\label{r3}
	\bigg\langle \int\frac{\text{d}\beta}{2\pi \text{i}}\sum_{i}e^{\beta \left( E-E_i\right) } \bigg\rangle=\int_{0}^{\infty} \overline{\rho\left(E \right)}\int\frac{\text{d}\beta}{2\pi \text{i}}e^{\beta \left(E- E^{\prime}\right) } \text{d}E,
\end{align}
so, the averaged density of states is:
\begin{align}\label{r4}
	\overline{\rho\left(E \right)}=	\bigg\langle \sum_{i}\delta \left( E-E_i\right)  \bigg\rangle.
\end{align}
On the other hand, using relations \eqref{a} and \eqref{amp}, the partition function of JT gravity can be expressed in terms of geodesic length states as follows:
\begin{align}\label{r5}
	\bigg\langle \sum_{i}e^{ -\beta E_i }\bigg\rangle=\int_{-\infty}^{\infty}	\langle\tau|\ell_1\rangle\overline{\langle\ell_1|\ell\rangle}\langle \ell| \beta-\tau\rangle\text{d}\ell_1\text{d}\ell,
\end{align}
where the average inner product $\overline{\langle\ell_1|\ell\rangle}$ is:
\begin{imageeq}
	\begin{overpic}[width=0.66\textwidth]{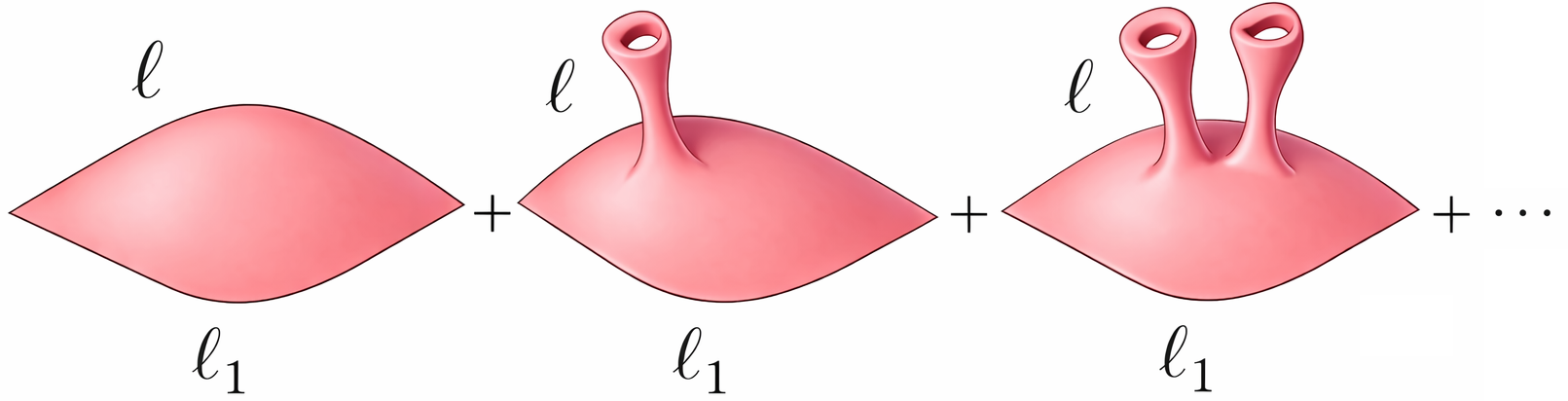}
		\put (100,11.2) {$\displaystyle ,$}
		\put (-14,11) {$\displaystyle \overline{\langle\ell_1|\ell\rangle}=$}            
	\end{overpic}
\end{imageeq}
and in the basis of energy, this can be written as:
\begin{align}\label{r6}
	\overline{\langle\ell_1|\ell\rangle}=\int_{0}^{\infty}\langle\ell_1|E\rangle \langle E|\ell\rangle\overline{\rho\left(E \right)} \text{d}E.
\end{align}
By substituting \eqref{r6} into \eqref{r5}, it becomes evident that \eqref{r1} can be derived. If we  replace $\overline{\rho\left(E \right)}$ from \eqref{r4} into \eqref{r6}, we get:
\begin{align}\label{r7}
	\overline{\langle\ell_1|\ell\rangle}=\int_{0}^{\infty}\langle\ell_1|E\rangle \langle E|\ell\rangle	\bigg\langle \sum_{i}\delta \left( E-E_i\right)  \bigg\rangle \text{d}E,
\end{align}
and by bringing all terms inside the angle brackets, we find:
\begin{align}\label{r9}
	\overline{\langle\ell_1|\ell\rangle}&=\bigg\langle\int_{0}^{\infty}\langle\ell_1|E\rangle \langle E|\ell\rangle	 \sum_{i}\delta \left( E-E_i\right)   \text{d}E\bigg\rangle\nonumber\\
	&=\bigg\langle\sum_{i}\langle\ell_1|E_i\rangle \langle E_i|\ell\rangle	    \bigg\rangle.
\end{align}
Guided by the above expression, Ref.~\cite{Iliesiu:2024cnh} proposed that the non-perturbative bulk inner product is given by:
\begin{align}\label{r10}
	\langle\ell_1|\ell\rangle=\sum_{i}\langle\ell_1|E_i\rangle \langle E_i|\ell\rangle.  
\end{align}
Recall that $\langle E_i | \ell \rangle$ is the scattering wavefunction introduced in \eqref{d}, denoted by $\phi_{E_i}(\ell)$ in Ref.~\cite{Iliesiu:2024cnh}. We have omitted the factor $e^{-S_0}$ that appears in the definition of the inner product in that reference. The inner product defined in this way satisfies the usual properties of conjugation and linearity,
and positivity of inner products. In this model, there are null states that are not physical.
The basis $|\ell\rangle$ is overcomplete. Nevertheless, despite this overcompleteness, once higher-genus
contributions are included, the path integral yields a well-defined inner product. 
If we approximate the sum $\sum_{i}$ by an integral weighted by the smooth density of states, $\int\rho_{0}(E)\, \text{d}E$, we  recover relation \eqref{2s10}.

In section \ref{without} we rewrite the computations of \cite{Stanford:2022fdt} by replacing the perturbative inner product \eqref{2s10} with the non-perturbative inner product \eqref{r10}. 

\section{The non-perturbative bulk Hilbert space with a baby universe}\label{4}
In this section, the inner product \eqref{r10} is generalized to geometries containing a single baby universe. To this end, an operator is first identified which, when inserted into the dual quantum mechanical ensemble, corresponds to the presence of a single geodesic boundary on the gravity side. Using this correspondence, the non-perturbative inner product in the geodesic-length basis is then introduced in the presence of a single baby universe, following the same procedure reviewed in the previous section.
\subsection{Operator insertion for a single geodesic boundary}
The full partition function of JT gravity with two
asymptotic boundaries  of regularized length $\beta$ and $\beta_1$, including contributions from all geometries, can be expressed as:
\begin{equation}\label{4s5}
	Z(\beta,\beta_1)=\int_{0}^{\infty}\mathcal{Z}	(\beta,a)Z_{\text{tr}}(\beta_1,a)a\text{d}a,
\end{equation}
where
\begin{equation}\label{1}
	\mathcal{Z}	(\beta,a)=\sum_{g=0}^{\infty}\mathcal{Z}_{g}	(\beta,a),
\end{equation}
and the definition of $\mathcal{Z}_{g}	(\beta,a)$ is given by \eqref{31}.
We want to derive 
$\mathcal{Z}	(\beta,a)$ non-perturbatively as the partition function of a dual quantum mechanics.
First, consider the following relation:
\begin{equation}\label{4s3}
	Z(\beta,\beta_1)= \left\langle \text{Tr} \, e^{-\beta H} \text{Tr}e^{-\beta_1H}\right\rangle- \left\langle \text{Tr} \, e^{-\beta H} \right\rangle \left\langle \text{Tr} \, e^{-\beta_{1} H} \right\rangle,
\end{equation}
Note that $	Z(\beta, \beta_1)$ is computed for connected geometries; therefore, the second term on the right-hand side of \eqref{4s3} is subtracted to remove contributions from disconnected geometries, which are already included in the first term.
By multiplying both sides of \eqref{4s5} by $\sqrt{4\pi \beta_1}\, e^{\frac{b^2}{4\beta_1}}$ and substituting $Z_{\text{tr}}(\beta_1,a)$ from \eqref{trup}, then performing the change of variables $\beta_1 \rightarrow \frac{1}{\mathrm{i}T}$ and integrating over $T$, one obtains:
\begin{equation}\label{9}
	\int_{-\infty}^{\infty}	Z(\beta,\frac{1}{\text{i} T})\sqrt{\frac{4\pi}{\text{i} T}}e^{\frac{\text{i} Tb^2}{4}}\text{d}T=\int_{0}^{\infty}\mathcal{Z}	(\beta,a)\int_{-\infty}^{\infty}e^{\frac{-\text{i}T}{4}(a^2-b^2)}\text{d}Ta\text{d}a.
\end{equation}
Integrating over $T$ on the right-hand side yields a delta function, and the subsequent integration over $a$ produces $\mathcal{Z}(\beta,a)$. This can be expressed more explicitly as follows:
\begin{equation}\label{12}
	\mathcal{Z}	(\beta,a)=	\int_{-\infty}^{\infty}	Z(\beta,\frac{1}{\text{i}4 T})\frac{e^{\text{i} Ta^2}}{\sqrt{\text{i}\pi T}}\text{d}T.
\end{equation}
By substituting $Z(\beta,\frac{1}{\text{i}4 T})$ from \eqref{4s3} into \eqref{12} and performing the integrals we obtain:
\begin{align}\label{10}
	\mathcal{Z}(\beta,a) 
	&= \int_{-\infty}^{\infty} \left\langle \text{Tr} \, e^{-\beta H} \, \text{Tr} \, e^{-\frac{H}{\mathrm{i}4T}} \right\rangle 
	\frac{e^{\mathrm{i} T a^2}}{\sqrt{\mathrm{i}\pi T}}  \mathrm{d}T 
	- \left\langle \text{Tr} \, e^{-\beta H} \right\rangle 
	\int_{-\infty}^{\infty} \left\langle \text{Tr} \, e^{-\frac{H}{\mathrm{i}4T}} \right\rangle 
	\frac{e^{\mathrm{i} T a^2}}{\sqrt{\mathrm{i}\pi T}}  \mathrm{d}T \nonumber \\
	&= \left\langle \text{Tr} \, e^{-\beta H} \, \frac{2}{a} \, \text{Tr} \, \cos\left(a H^{\frac{1}{2}}\right) \right\rangle 
	- \left\langle \text{Tr} \, e^{-\beta H} \right\rangle 
	\int \overline{\rho(E)} \, e^{-\frac{E}{\mathrm{i}4T}} 
	\frac{e^{\mathrm{i} T a^2}}{\sqrt{\mathrm{i}\pi T}}  \mathrm{d}T  \mathrm{d}E \nonumber \\
	&= \left\langle \text{Tr} \, e^{-\beta H} \, \frac{2}{a} \, \text{Tr} \, \cos\left(a H^{\frac{1}{2}}\right) \right\rangle 
	- \left\langle \text{Tr} \, e^{-\beta H} \right\rangle 
	\int_{0}^{\infty} \overline{\rho(E)} \, \frac{2}{a} \cos\left(a E^{\frac{1}{2}}\right)  \mathrm{d}E.
\end{align}
In the second line of the above expression, relation \eqref{r1p} was used. For simplicity, define the following operator:
\begin{equation}\label{01}
	\mathcal{O}\left( a\right)= \frac{2}{a}\text{Tr}\cos\left( a H^{\frac{1}{2}}\right)-\frac{2}{a}\int  \overline{\rho\left(E \right)} \cos\left( a E^{\frac{1}{2}}\right)\text{d}E.
\end{equation}
With this definition, the expression \eqref{10} can be rewritten as:
\begin{equation}\label{tatalo}
	\mathcal{Z}	(\beta,a)=\left\langle \text{Tr} \, e^{-\beta H}  \mathcal{O}\left( a\right)\right\rangle.
\end{equation}
We aim to express the second term in \eqref{01} using random matrix theory so as to define an operator entirely within the random-matrix framework. In this form, the right-hand side of the above relation provides a direct manifestation of the duality. To this end, the equation of motion obtained by varying the action in \eqref{zm} with respect to $E_i$ is used:
\begin{align}\label{taslo}
LV^{\prime}\left( E_i\right)&= 2\sum_{i\neq j}^{L}\frac{1}{E_i-E_j}
\nonumber\\&=2\int\sum_{j}^{L}\delta\left(E-E_j \right)\frac{1}{E_i-E}\text{d}E. 
\end{align}
Equivalently,
\begin{align}\label{tas}
	LV\left( -z^2\right)=2\int\sum_{j}^{L}\delta\left(E-E_j \right)\log\left( -z^2-E\right) \text{d}E,
\end{align}
 where $E_i=-z^2$. Averaging both sides of the above relation and using \eqref{r4} gives
 \begin{align}\label{tzas}
 	LV\left( -z^2\right)=2\int\overline{\rho\left(E \right)}\log\left( -z^2-E\right) \text{d}E,
 \end{align} 
Multiplying by \(e^{az}\) and integrating over \(z\) along an inverse Laplace transform contour yields:
 \begin{equation}\label{0x1}
 \frac{L}{2}W\left(a \right)= \frac{2}{a}\int  \overline{\rho\left(E \right)} \cos\left( a E^{\frac{1}{2}}\right)\text{d}E.
 \end{equation}
where $W(a)$ denotes the inverse Laplace transform of the potential,
 \begin{align}\label{L}
 	W\left(a \right) =\frac{1}{2\pi \text{i}}\int_{-\text{i}\infty}^{\text{i}\infty}V\left(-z^2 \right)e^{az} \text{d}z.
 \end{align}
 As a result, the operator in \eqref{01} takes the form
\begin{equation}\label{02}
	\mathcal{O}\left( a\right)= \frac{2}{a}\sum_{i=1}^{L}\cos\left( a E_{i}^{1/2}\right)-\frac{L}{2}W\left(a \right),
\end{equation}
Therefore, JT gravity path integral with a single geodesic boundary is dual to inserting the above operator into the dual quantum mechanics ensemble \cite{Goel:2020yxl,Blommaert:2021fob}. The operator \eqref{02} is the inverse Laplace transform of the operator that is dual to a single FZZT boundary:
\begin{equation}\label{03}
	\mathcal{O}_{\text{FZZT}}\left( E\right)=
	\text{Tr}\log\left(E-H \right)-\frac{L}{2} V\left(E \right),
\end{equation}
where the first term, $\text{Tr}\log\left(E-H \right)$, plays the role of imposing a fixed-energy boundary condition in the matrix model (instead of fixed temperature), and where FZZT branes,\footnote{FZZT branes have been widely studied in \cite{ Goel:2021wim,Fateev:2000ik,Ponsot:2001ng,Mertens:2020hbs,Hosomichi:2008th,Teschner:2000md}.} being exponentials of boundaries, correspond in the matrix ensemble to \cite{Saad:2019lba, Okuyama:2021eju, Maldacena:2004sn, Blommaert:2019wfy}:
\begin{equation}\label{04}
	\psi_{\text{FZZT}}\left(E\right)=\det\left(E-H \right)\exp\left( -\frac{L}{2} V\left(E \right)\right).
\end{equation}
The determinant operators serve to capture non-perturbative effects.

\subsection{Non-perturbative geodesic inner product}

According to expression \eqref{a1}, $\mathcal{Z}(\beta, a)$ can be rewritten in the geodesic length basis as:
\begin{equation}\label{11}
	\mathcal{Z}	(\beta,a)=	\int_{-\infty}^{\infty}\text{d}\ell_1\text{d}\ell	\langle\tau|\ell_1\rangle\overline{\langle\ell_1|\ell,a\rangle}\langle \ell| \beta-\tau\rangle,
\end{equation}
where
\begin{imageeq}
	\begin{overpic}
		[width=0.66\textwidth]{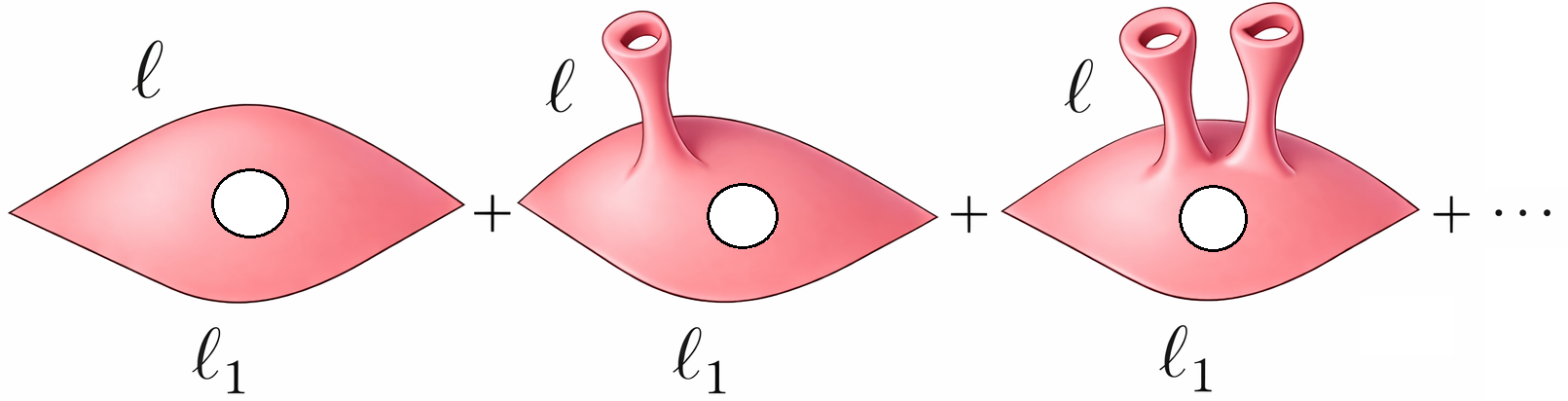}
		\put (-17,11) {$\displaystyle \overline{\langle\ell_1|\ell,a\rangle}=$}	
		\put (50.5,11) {$\displaystyle a$}
		\put (19,11) {$\displaystyle a$}
		\put (81,11) {$\displaystyle a$}			
	\end{overpic}
	\label{1u}	
\end{imageeq}
Note that the overline indicates that this geodesic inner product is non-perturbative and includes contributions from all genera. The left-hand side of \eqref{11} can now be replaced using the dual quantum mechanical ensemble defined in \eqref{tatalo}:
\begin{equation}\label{13}
	\left\langle \text{Tr} \, e^{-\beta H}  \mathcal{O}\left( a\right)\right\rangle=	\int_{-\infty}^{\infty}\text{d}\ell_1\text{d}\ell	\langle\tau|\ell_1\rangle\overline{\langle\ell_1|\ell,a\rangle}\langle \ell| \beta-\tau\rangle
\end{equation}
In terms of  the density of eigenvalues, the above relation can be expressed as:
\begin{equation}\label{4s13}
	\left\langle \text{Tr} \, e^{-\beta H}  \mathcal{O}\left( a\right)\right\rangle=	\int_{0}^{\infty} \overline{\rho\left(E,a \right)} e^{-\beta E}\text{d}E,
\end{equation}
where
\begin{align}\label{sr2}
	\overline{\rho\left(E,a \right)}=\sum_{g=0}^{\infty} e^{-2g  S_0}\rho_{g}\left(E ,a\right),
\end{align}
and $\rho_{g}\left(E ,a\right)$ defined by \eqref{density}. Applying the inverse Laplace transform to \eqref{4s13} yields:
\begin{equation}\label{4s14}
	\overline{\rho\left(E,a \right)}=	\bigg\langle \sum_i\delta\left(E-E_i \right) 	\mathcal{O}\left( a\right)\bigg\rangle.	 
\end{equation}
It is evident that substituting
\begin{equation}\label{4s15}
	\overline{\langle\ell_1|\ell,a\rangle}=	\int_{0}^{\infty}	\overline{\rho\left(E,a \right)}\langle\ell_1|E\rangle \langle  E| \ell\rangle\text{d}E,
\end{equation}
into equation \eqref{13} reproduces expression \eqref{4s13}.
In the above expression, $\overline{\rho\left(E,a \right)}$ can be replaced using \eqref{4s14}; performing the integral then yields:
\begin{align}\label{4s16}
	\overline{\langle\ell_1|\ell,a\rangle}&=	\int_{0}^{\infty}\text{d}E	\bigg\langle \sum_i\delta\left(E-E_i \right) 	\mathcal{O}\left( a\right)\bigg\rangle\langle\ell_1|E\rangle \langle  E| \ell\rangle\nonumber\\
	&=		\bigg\langle \sum_i\langle\ell_1|E_i\rangle \langle  E_i| \ell\rangle	 	\mathcal{O}\left( a\right)\bigg\rangle.
\end{align}
From the above equation, and by analogy with the approach used to define the inner product in \eqref{r10}, one can suggest that:
\begin{align}\label{4s17}
	\langle\ell_1|\ell,a\rangle=\sum_{i}\mathcal{O}\left( a\right)\langle\ell_1|E_i\rangle \langle E_i|\ell\rangle.
\end{align}
Note that, if the sum $\sum_i\mathcal{O}\left( a\right)$ is approximated by the integral $\int \rho_{\text{tr}}(E,a)\, \mathrm{d}E$ using the trumpet density of states, we obtain:
{	\begin{equation}\label{20}
		\langle\ell_1|\ell,a\rangle=\int \rho_{\text{tr}}(E,a) \langle\ell_1|E\rangle	\langle E| \ell\rangle\text{d}E.
\end{equation}}
	
	Consider a general superposition of geodesic states with wave function $\psi_{a}\left( \ell\right) $:
	\begin{equation}
		|\psi\left(a \right) \rangle=\int  \psi_{a}\left( \ell\right) 	|\ell,a\rangle\text{d}\ell.
	\end{equation}
	The properties of the inner product between such states become transparent upon expressing
	$\psi_{a}\left( \ell\right) $ as a superposition of scattering wave functions $	\langle E|\ell\rangle$ via the expansion:
	\begin{equation}
		\hat{\psi}_{a}\left(E \right) :=\int \psi_{a}\left( \ell\right) 	\langle E|\ell\rangle\text{d}\ell .
	\end{equation}
Using \eqref{2s10}, one can also invert this relation and write:
	\begin{equation}
		\psi_{a}\left( \ell\right) =\int \text{d}E	\rho_{0}\left(E \right) \langle \ell|E\rangle	\hat{\psi}_{a}\left(E \right).  
	\end{equation}
	The positivity of the inner product \eqref{4s17} is evident from the following expression: 
	\begin{align}
		\langle	\psi\left(a \right)|\psi\left(a \right) \rangle=\int \psi_{a}\left( \ell\right)  \psi_{a}^{\ast}\left( \ell^{\prime}\right) \langle \ell^{\prime},a	|\ell,a\rangle\text{d}\ell\text{d}\ell^{\prime}&=		\sum_{k}\int \psi_{a}\left( \ell\right)  \psi_{a}^{\ast}\left( \ell^{\prime}\right)\langle \ell^{\prime}|E_k\rangle	\langle E_k| \ell\rangle \mathcal{O}^{2}\left( a\right)\text{d}\ell\text{d}\ell^{\prime}\nonumber\\& =
		\sum_{k}|	\hat{\psi}_{a}\left(E_{k} \right)|^{2} \mathcal{O}^{2}\left( a\right),
	\end{align}
	which is manifestly non-negative.
		A notable feature of the gravitational Hilbert space is the presence of redundancy in the form of null states--i.e., states for which $\hat{\psi}_{a}(E_k) = 0$ for all $E_k$. To obtain a well-defined Hilbert space, such null states must be excluded from the vector space. 
\section{Wormhole without  emission of a baby universe}\label{without}
Using the inner product \eqref{r10}, the probability of finding a wormhole of length $\ell$ can be computed. The probability distribution for a wormhole of length $\ell$, in the absence of baby-universe emission, is expressed as:
\begin{equation}\label{wbet}
	\hat{P}_{\beta}\left( \ell\right)= \int_{-\infty}^{\infty}\text{d}\ell_{1}\text{d}\ell_{2}\langle\frac{\beta}{2}+\text{i}t|\ell_{2}\rangle\langle\ell_{2}|\ell\rangle\langle\ell|\ell_{1}\rangle\langle\ell_{1}|\frac{\beta}{2}+\text{i}t\rangle.
\end{equation}
The notation $\hat{P}$ represents an unnormalized probability distribution. To achieve proper normalization, we must divide $\hat{P}$ by the partition function. The inverse Laplace transformation yields the fixed-energy version of the probability distribution as follows:
\begin{align}
	\hat{P}\left( \ell\right)=\sum_{k,j}\int\frac{\text{d}\beta}{2\pi \text{i}}e^{\beta E} 	\int_{-\infty}^{\infty}\text{d}\ell_{1}\text{d}\ell_{2}\langle\frac{\beta}{2}+\text{i}t|\ell_2\rangle\langle\ell_2|E_k\rangle\langle E_k| \ell\rangle 
	\langle \ell| E_{j}\rangle \langle E_{j}|\ell_1\rangle\langle\ell_1|\frac{\beta}{2}+\text{i}t\rangle  .
\end{align}
Note that the inner product \eqref{r10} has been used in the above expression.
By substituting the inner product $ \langle E|\frac{\beta}{2}+\text{i}t\rangle\nonumber=e^{(-\beta/2+\text{i}t)E} $ in the above relation, we have: 
\begin{align}\label{rob3}
	\hat{P}\left( \ell\right)=\sum_{k,j}\int\frac{\text{d}\beta}{2\pi \text{i}}e^{\beta \left( E-\frac{E_k+E_{j}}{2}\right) }	~e^{-\text{i}t\left( E_k-E_{j}\right) }\langle E_{k}|\ell\rangle\langle\ell|E_{j}\rangle.
\end{align}
The integral over $ \beta $ sets $ E_k + E_{j} = 2E $.  Performing the change of variables:
\begin{equation}\label{chae}
	E_k=E+\sqrt{E}\omega_{k} \qquad E_{j}=E-\sqrt{E}\omega_{k},  
\end{equation}
in \eqref{rob3}, we obtain:
\begin{align}\label{1op}
	\hat{P}\left( \ell\right)&= \sum_{\omega_{k}} 	~e^{-\text{i}2 \sqrt{E}t\omega_{k} }\langle E-\sqrt{E}\omega_{k}|\ell\rangle\langle\ell|E+\sqrt{E}\omega_{k}\rangle.
\end{align}
The parameter $\omega_{k}$ ranges over both positive and negative values. In the semiclassical regime, where $E$ is large, the wave function $\langle E|\ell\rangle$ can be approximated as:
\begin{equation} \label{vlt}
	\langle E|\ell\rangle\approx\frac{\left( 8\pi\right) ^{1/2}}{E^{1/4}}e^{-\pi \sqrt{E}}\cos \left( \sqrt{E}\left(\ell +\log\left(4 E \right)-2 \right) -\frac{\pi}{4}\right) .
\end{equation}
In the semi-classical regime, the probability distribution for a wormhole of length 
$\ell$ without baby universe emission takes the form:
\begin{align}\label{114r}
	\hat{P}\left( \ell\right)&\approx \frac{ 2\pi e^{-2\pi \sqrt{E}} }{E^{1/2}}\sum_{\omega_{k}} 	e^{-\text{i}\ell_t\omega_k }\left( e^{\pm\text{i}\ell\omega_{k} }+ e^{\pm\text{i}2\ell \sqrt{E} }\right),
\end{align}
where $\ell_t =2\sqrt{E}t$. Notice unimportant $ \log(4E) $ term has been omitted.
Similar to equation (2.15) of \cite{Stanford:2022fdt}, the expression above contains four terms. Upon approximating the discrete sum by the continuous disk spectral density,
\(\sum_{i,j} \rightarrow \int \rho_{0}(E_1)\,\rho_{0}(E_2)\,\mathrm{d}E_1\,\mathrm{d}E_2\),
one recovers equation (2.17) of that reference. Within this approximation, only the expanding branch--namely, the term \(e^{-\mathrm{i}\omega_k(\ell_t - \ell)}\)--survives. 
It is also evident from \eqref{114r} that, in the non-perturbative regime, the dominant contribution continues to arise from $e^{-\text{i}\omega_{k}(\ell - \ell_t)}$, while $e^{-\text{i}\omega_{k}(\ell + \ell_t)}$ and $e^{-\text{i}\omega_{k}\ell_t} e^{\pm \text{i}2\ell \sqrt{E}}$ provide no significant contribution.


\section{Wormhole with emission of a baby universe}\label{6}
In the presence of a single baby universe, the geometry includes an additional geodesic boundary. This modifies the gravitational path integral and its dual description in the random matrix theory. As discussed earlier, the insertion of the operator $\mathcal{O}(a)$ into the matrix ensemble corresponds to such a configuration. Consequently, the geodesic inner product becomes non-perturbative and depends on the additional parameter $a$, which is conjugate to the length of the emitted baby universe. The fixed-energy probability mass associated with baby universe emission is therefore given by:
\begin{equation}\label{prob2}
	\hat{P}_{1}\left( \ell\right)=\int\frac{\text{d}\beta}{2\pi \text{i}}e^{\beta E} \int_{-\infty}^{\infty}\text{d}\ell_{1}\text{d}\ell_{2}\int \text{d}a\text{d}s\langle\frac{\beta}{2}+\text{i}t|\ell_{2}\rangle\langle\ell_{2}|\ell,a\rangle\langle\ell,a|\ell_{1}\rangle\langle\ell_{1}|\frac{\beta}{2}+\text{i}t\rangle.
\end{equation}
Here, the subscript $1$ denotes the emission of a single baby universe. Following the same procedure as in the no-baby-universe case, we insert the non-perturbative geodesic inner product \eqref{4s17} into \eqref{prob2} and perform the integrals over $\ell_1$ and $\ell_2$, obtaining:
\begin{align}\label{prob3}
	\hat{P}_{1}\left( \ell\right)=\sum_{k,j}\int\frac{\text{d}\beta}{2\pi \text{i}}e^{\beta E} \int \text{d}a\text{d}s	\langle\frac{\beta}{2}+\text{i}t|E_k\rangle\langle E_k| \ell\rangle 
	\langle \ell| E_{j}\rangle \langle E_{j}|\frac{\beta}{2}+\text{i}t\rangle \mathcal{O}\left( a\right)\mathcal{O}\left( a\right).
\end{align}
After substituting the inner product $\langle E|\frac{\beta}{2}+\text{i}t\rangle = e^{(-\beta/2+\text{i}t)E}$, performing the $\beta$ integral which imposes the condition $E_k + E_j = 2E$ and applying the change of variables \eqref{chae}, the probability distribution \eqref{prob3} simplifies in the semi-classical regime to: 
\begin{align}\label{114}
	\hat{P}_{1}\left( \ell\right)\approx \frac{ 2\pi e^{-2\pi \sqrt{E}}}{E^{1/2}}\sum_{\omega_{k}} 	e^{-\text{i}\ell_t\omega_{k} }\left( e^{\pm\text{i}\ell\omega_{k} }+ e^{\pm\text{i}2\ell \sqrt{E} }\right)	
	\int \text{d}a\text{d}s~\mathcal{O}\left( a\right)\mathcal{O}\left( a\right). 
\end{align}
Unlike relation (3.9) in \cite{Stanford:2022fdt}, expression \eqref{114} exhibits no apparent connection between the baby-universe length $a$, the black hole age $\ell_t$, and the wormhole length $\ell$. At genus one, the expanding term $e^{\mathrm{i}(\ell + a - \ell_t)\omega}$ is interpreted as corresponding to a smooth geometry, whereas the contracting term $e^{\mathrm{i}(-\ell + a - \ell_t)\omega}$ is associated with a firewall geometry.
 Such distinctions do not appear in the present framework, as expected, since the inner product \eqref{4s17} indicates that the baby-universe length is independent of the wormhole lengths. Nevertheless, relation \eqref{20} suggests an implicit dependence between \(a\) and \(\ell\), as the presence of a baby universe modifies the discrete energy eigenvalue distribution. More specifically, the sum \(\sum_i \mathcal{O}(a)\) is approximated by the integral \(\int \rho_{\text{tr}}(E,a)\,\mathrm{d}E\), which captures the effect of the baby universe on the spectral density. However, the explicit relation between \(a\), \(\ell\), and \(\ell_t\) in the non-perturbative regime remains unclear. In the following, different geometric structures are examined, guided by methods inspired by the perturbative regime.
  

\subsection{Smooth geometry}\label{i}
The probability of finding a smooth geometry following the emission of a baby universe is now computed. A crucial step in this calculation is determining the integration ranges for the twist parameter and geodesic length. Using geometric considerations, Ref.~\cite{Stanford:2022fdt} introduced a no-shortcut condition, which is assumed to continue holding. Under this condition, the twist parameter for a smooth geometry ranges as $0 < s < a$, while the geodesic boundary satisfies $0 < a < \ell_t$. Accordingly, the probability is given by:
\begin{align}\label{115}
	\hat{P}_{\text{smooth}}&\approx \frac{ 2\pi e^{-2\pi \sqrt{E}} }{E^{1/2}} \int_{\delta}^{\ell_t} \text{d}a\int_{0}^{a}\text{d}s~
	\mathcal{O}\left( a\right)\mathcal{O}\left( a\right)\\&
	=-\frac{ 2\pi e^{-2\pi \sqrt{E}} }{E^{1/2}}\sum_{i,j}\left(  F(E_i, E_j, \ell_t)+ 4\gamma + 4\log\delta\right) ,	
\end{align}
where the definition of $F(E_i, E_j, \ell_t)$ is given in \eqref{P}. The details of the calculation of the above integral are presented in Appendix \ref{apen}. For the limit $\ell_t \rightarrow \infty$, the function $F(E_i, E_j, \ell_t)$ reduces to:
\begin{align}\label{pP}
	F(E_i, E_j) =
	\begin{cases}
		\begin{aligned}[t]
			&2\log |E_i - E_j| ,
		\end{aligned}
		& E_i \neq E_j, \\[8pt]
		\begin{aligned}[t]
			&\log \left(4 E_i \epsilon^2\right) ,
		\end{aligned}
		& E_i = E_j,
	\end{cases}
\end{align}
so that
\begin{align}\label{11w}
	\hat{P}_{\text{smooth}}=\text{constant}.
\end{align}

\subsection{Firewall geometry  }\label{three}
As shown in Ref.~\cite{Stanford:2022fdt}, for genus one in the presence of a firewall, the no-shortcut condition constrains the twist parameter $s$ to
$
a - \ell_t < s < \ell_t,
$
and restricts the geodesic boundary length to
$
\ell_t < a < 2 \ell_t.
$
Assuming this condition holds in the present calculation, the probability of encountering a geometry with a firewall following the emission of a large baby universe is given by
\begin{align}\label{116}
	\hat{P}_{\text{firewall}}&\approx \frac{2\pi e^{-2\pi \sqrt{E}}}{E^{1/2}}
	\int_{\ell_t}^{2\ell_t} \text{d}a\int_{a-\ell_t}^{\ell_t}\text{d}s~ 
	\mathcal{O}\left( a\right)\mathcal{O}\left( a\right)\\&\approx\frac{ 2\pi e^{-2\pi \sqrt{E}} }{E^{1/2}}
	\sum_{i, j}  	\left( 2\ell_t	Q_2(E_i,E_j,\ell_t)-Q_1(E_i,E_j,\ell_t)\right).
\end{align}
where the functions $Q_1(E_i, E_j, \ell_t)$ and $Q_2(E_i, E_j, \ell_t)$ are defined in \eqref{od1} and \eqref{odh1}, respectively. In the limit $\ell_t \to \infty$, these functions reduce to
\begin{gather}
	Q_1(E_i, E_j, \ell_t) \simeq \log 4, \qquad \text{for } E_i = E_j, \\
	Q_2(E_i, E_j, \ell_t) \simeq \ell_t^{-1}, \qquad \text{for } E_i = E_j.
\end{gather}
Consequently, at very late time one finds:
\begin{align}
	\hat{P}_{\text{firewall}} &\approx \frac{2\pi e^{-2\pi \sqrt{E}}}{E^{1/2}} 
	\sum_{i, j} (2 - 2\log 2), \label{1ir}
\end{align}

\section{Remarks} \label{7}

This work investigated wormhole configurations in JT gravity within a non-perturbative bulk Hilbert space framework, with particular emphasis on the role of baby universes in mediating the tunneling of an old black hole into a white hole and the emergence of a firewall at the horizon at late times. It was shown that the probabilities of encountering smooth geometries and firewall geometries at late times approach constant values, as expected from the late-time (plateau time $t \sim e^{S(E)}$ )  behavior of the spectral form factor (SFF) in systems without time-reversal symmetry. This follows from the fact that the normalized SFF  
\begin{align}
 |\langle\text{TFD}\left(t^{\prime} \right) |\text{TFD}\left(t \right)\rangle|^{2}=\frac{Z\left(\beta+ \text{i}\left(t-t^{\prime} \right)  \right)Z\left(\beta-\text{i}\left(t-t^{\prime} \right)  \right) }{Z\left(\beta\right)^{2} }
\end{align}
is the projection of the TFD state at time $t$ with one at time $t^{\prime}$.
In this work, the overlap of the late-time TFD state with states of bulk age 
$t^{\prime}$, defined with respect to a special spatial slice, was computed. Then, the very late-time behavior of this calculation was considered to examine smooth and firewall geometries. 
It was conjectured that at very late times 
$P_{\text{smooth}}\left(\infty \right) =P_{\text{firewall}}\left(\infty \right)=\frac{1}{2}$
and that the physics of the plateau could be relevant \cite{Stanford:2022fdt}. In the present analysis, the saturation of these probabilities is observed, but their equality is not realized.
It is worth noting that the results in \cite{Zolfi:2024ldx} do not exhibit equality between 
$P_{\text{smooth}}\left(t\right) $ and 
$P_{\text{firewall}}\left(t \right)$ for genus-two geometries or for scenarios involving the emission of multiple baby universes, although their time-dependent behavior is the same.

As noted earlier, the explicit relationship between the size of the baby universe, the wormhole length, and the black hole age remains unclear in the non-perturbative regime. In our calculation, we have assumed that the no-shortcut condition—essential in the perturbative regime—also applies in this non-perturbative context. An interesting open question is the precise nature of the no-shortcut condition in the non-perturbative regime and its meaning within the dual random matrix theory.
\section*{Acknowledgments}
We thank  Mohsen Alishahiha, Ahmad Poostforush and Amir Abbas Khodahami  for
useful discussions. 
Additionally, OpenAI's ChatGPT (GPT-5) was used to refine the text and assist in producing 3D-style coloring for figures.

\appendix

\section{Details of Calculations}\label{apen}
To evaluate the integral in \eqref{115}, the definition in \eqref{02} is employed, giving:
\begin{align}\label{O}
	\int_{\delta}^{\ell_t} \mathcal{O}(a)\mathcal{O}(a) \, a\mathrm{d}a 
	&= \sum_{i,j=1}^{L} 4 \int_{\delta}^{\ell_t} \frac{\mathrm{d}a}{a} \, e^{-\epsilon a} \cos\left(a\sqrt{E_i}\right) \cos\left(a\sqrt{E_j}\right) \nonumber \\
	&\quad - 2L \sum_{i=1}^{L} \int_{\delta}^{\ell_t} W(a) e^{-\epsilon a} \cos\left(a\sqrt{E_i}\right) a\mathrm{d}a \nonumber \\
	&\quad + \frac{L^2}{4} \int_{\delta}^{\ell_t} W(a) W(a) \, a\mathrm{d}a.
\end{align}
The first term in expression \eqref{O} is evaluated with the following relation:
\begin{align}\label{ope1}
	-4\int_{\delta}^{\ell_t} \frac{\mathrm{d}a}{a} e^{-\epsilon a} \cos\left(a\sqrt{E_i}\right) \cos\left(a\sqrt{E_j}\right) = F(E_i, E_j, \ell_t) + 4\gamma + 4\log\delta,
\end{align}
where, for $\ell_t \gg 1$, the function $F(E_i, E_j, \ell_t)$   is defined as:
\begin{align}\label{P}
	F(E_i, E_j, \ell_t) =
	\begin{cases}
		\begin{aligned}[t]
			&2\log |E_i - E_j| - \frac{2 \sin\left((\sqrt{E_i} + \sqrt{E_j}) \ell_t\right)}{(\sqrt{E_i} + \sqrt{E_j}) \ell_t} - \frac{2 \sin\left((\sqrt{E_i} - \sqrt{E_j}) \ell_t\right)}{(\sqrt{E_i} - \sqrt{E_j}) \ell_t},
		\end{aligned}
		& E_i \neq E_j, \\[8pt]
		\begin{aligned}[t]
			&\log \left(4 E_i \epsilon^2\right) - \frac{\sin \left(2 \sqrt{E_i} \ell_t\right)}{\sqrt{E_i} \ell_t}+ \dots,
		\end{aligned}
		& E_i = E_j.
	\end{cases}
\end{align}
 Using \eqref{03} and performing the integration over 
$a$, the second term becomes:
\begin{align}
	\int_{\delta} ^{\ell_t} W\left( a\right) e^{-\epsilon a}\cos\left(a\sqrt{E_i} \right)a\text{d}a=-\underbrace{\int_{-\infty}^{\infty}\frac{V\left(z^2\right) }{2
			\pi  z^2}\text{d}z}_{\text{diverges}}+\underbrace{\int_{-\infty}^{\infty}\frac{V\left(z^2\right) e^{\text{i} \ell_t   z} (1-\text{i} \ell_t  z)}{2
			\pi  z^2}\text{d}z}_{0}.
\end{align}
By applying the same procedure to the third term, we obtain:
\begin{align}\label{qq}
	\int_{\delta} ^{\ell_t} W\left( a\right) W\left( a\right) a\text{d}a=\ell_t\int _{\mathbb{R} }\frac{e^{\text{i}u}\left(1 - \text{i} u\right)-1  
	}{8\pi^2 u ^2}G(\frac{u}{\ell_t})\text{d}u,	
\end{align}
where
\begin{equation}
	G(u)=\int_{\mathbb{R} }V\left(\frac{(u+v)^{2}}{4} \right) \,V\left(\frac{(u-v)^{2}}{4} \right)\text{d}v.
\end{equation}
For large $\ell_t$,
the leading contribution from $G(u/\ell_t)$ to the integral \eqref{qq} is given by:
\begin{align}
	\ell_t G(0)\underbrace{\int _{\mathbb{R} }\frac{e^{\text{i}u}\left(1 - \text{i} u\right) 
		}{8\pi^2 u ^2}\text{d}u}_{0}
	+O\left(\frac{1}{\ell_t} \right).
\end{align}
Using the results above, expression \eqref{O} can be written as:
\begin{align}
	\int_{\delta}^{\ell_t} \mathcal{O}(a)\mathcal{O}(a) a\mathrm{d}a 
	&= -\sum_{i,j=1}^{L} F(E_i, E_j, \ell_t) + \text{constant}.
\end{align}
where the constant term on the second line is independent of $E_i$ and thus cancels upon normalizing the matrix integral.

The following integral from \eqref{116} is now evaluated:
\begin{align}
	\int_{\ell_t}^{2\ell_t} \mathcal{O}(a)\mathcal{O}(a) \left(2\ell_t-a\right) \mathrm{d}a. 
\end{align}
One can show that the following terms appearing in the above integral (after substituting \eqref{02}) all vanish:
\begin{align}
	\int_{\ell_t}^{2\ell_t} W(a) W(a) \mathrm{d}a
	= \frac{\mathrm{i}\ell_t}{8\pi^2} \int_{\mathbb{R}} \frac{e^{\mathrm{i} 2u} - e^{\mathrm{i} u}}{u} G\left(\frac{u}{\ell_t}\right) \mathrm{d}u = 0,
\end{align}
\vspace{-5mm}
\begin{align}\label{qqfr}
	\int_{\ell_t}^{2\ell_t} W(a) W(a) a \mathrm{d}a
	= \ell_t \int_{\mathbb{R}} \frac{e^{2\mathrm{i} u}(1 - 2\mathrm{i} u) + e^{\mathrm{i} u}(\mathrm{i} u - 1)}{8\pi^2 u^2} G\left(\frac{u}{\ell_t}\right) \mathrm{d}u = 0,
\end{align}
and
\begin{align}
	\int_{\ell_t}^{2\ell_t} W(a) e^{-\epsilon a} \cos\left(a\sqrt{E_i}\right) \mathrm{d}a = 0.
\end{align}
The only terms that survive come from the product of cosines. It is therefore more convenient to use the following relation:
\begin{equation}\label{od1}
	\begin{aligned}
		Q_1(E_i, E_j, \ell_t) &=4\int_{\ell_t}^{2\ell_t} \frac{\mathrm{d}a}{a} 
		\cos(a\sqrt{E_i}) \cos(a\sqrt{E_j}) \\
		&\simeq
		\begin{cases}
			\ell_t^{-1}	A_{ij}+\dots, 
			& E_i \neq E_j, \\\vspace{3mm}
			\displaystyle
			\log 4-\frac{4}{\ell_t  \sqrt{E_i}} \sin ^3\left(\ell_t  \sqrt{E_i}\right) \cos \left(\ell_t
			\sqrt{E_i}\right)+\dots, 
			& E_i = E_j.
		\end{cases}
	\end{aligned}
\end{equation}
where
\begin{align}
	A_{ij} =
	& \frac{\sin
		\left(2 \ell_t 
		\left(\sqrt{E_i}-\sqrt{E_j}\right)\right)}{\sqrt{E_i}-\sqrt{E_j}}-\frac{2 \sin \left(\ell_t
		\left(\sqrt{E_i}-\sqrt{E_j}\right)\right)}{\sqrt{E_i}-\sqrt{E_j}}\\\nonumber
	&-\frac{2
		\sin \left(\ell_t 
		\left(\sqrt{E_i}+\sqrt{E_j}\right)\right)}{\sqrt{E_i}+\sqrt{E_j}}+\frac{\sin
		\left(2 \ell_t
		\left(\sqrt{E_i}+\sqrt{E_j}\right)\right)}{\sqrt{E_i}+\sqrt{E_j}},
\end{align}
and
\begin{equation}\label{odh1}
	\begin{aligned}
		Q_2(E_i, E_j, \ell_t) &=4\int_{\ell_t}^{2\ell_t} \frac{\mathrm{d}a}{a^{2}} 
		\cos(a\sqrt{E_i}) \cos(a\sqrt{E_j}) \\
		&\simeq
		\begin{cases}
			\ell_t^{-2}	B_{ij}+\dots, 
			& E_i \neq E_j, \\\vspace{4mm}
			\displaystyle
			\ell_t^{-1}+	\frac{\sin \left(4 \ell_t  \sqrt{E_i}\right)-4 \sin \left(2 \ell_t
				\sqrt{E_i}\right)}{4 \ell_t^2 \sqrt{E_i}}+\dots, 
			& E_i = E_j,
		\end{cases}
	\end{aligned}
\end{equation}
where
\begin{align}
	B_{ij} = \frac{1}{E_i - E_j} \biggl[ 
	&\sqrt{E_i} \sin(\ell_t \sqrt{E_i}) \bigl( \cos(\ell_t (\sqrt{E_i} - 2\sqrt{E_j})) + \cos(\ell_t (\sqrt{E_i} + 2\sqrt{E_j})) - 4\cos(\ell_t \sqrt{E_j}) \bigr)\nonumber \\
	- &\sqrt{E_j} \sin(\ell_t \sqrt{E_j}) \bigl(\cos(\ell_t (2\sqrt{E_i} - \sqrt{E_j}))  + \cos(\ell_t (2\sqrt{E_i} + \sqrt{E_j})) \bigr) 
-4\cos(\ell_t \sqrt{E_i})	\biggr]
\end{align}
Note that the answers to the above integrals were approximated for  $\ell_t\gg1$.

\bibliographystyle{utphys.bst}
\bibliography{ref2.bib}

\end{document}